\documentclass[10pt,american,english]{article}
\usepackage[T1]{fontenc}
\usepackage[latin9]{inputenc}
\usepackage[letterpaper]{geometry}
\usepackage{float}
\usepackage{textcomp}
\usepackage{amstext}
\usepackage{amssymb}
\usepackage{graphicx}
\usepackage{setspace}
\usepackage[numbers]{natbib}

\makeatletter

\usepackage{multicol}
\usepackage{fancyhdr}
\usepackage{titlesec}
\usepackage[flushleft]{threeparttable}
\usepackage{caption}
\usepackage{float}



\titleformat*{\section}{\normalsize\bfseries}
\titleformat*{\subsection}{\normalsize\itshape}
\titleformat*{\subsubsection}{\normalsize\itshape}
\titleformat*{\paragraph}{\normalsize\itshape}
\titleformat*{\subparagraph}{\normalsize\itshape}

\makeatother

\usepackage{babel}
\begin{document}
\begin{center}
IAC-16-C1.6.7.32480
\par\end{center}

\medskip{}

\begin{center}
\textbf{ORBITAL DYNAMICS OF A SOLAR SAIL ACCELERATED BY THERMAL DESORPTION
OF COATINGS}
\par\end{center}

\medskip{}

\begin{center}
\textbf{Elena Ancona}\textsuperscript{a} and \textbf{Roman Ya. Kezerashvili}\textsuperscript{b}
\par\end{center}

\begin{center}
\textsuperscript{a}Politecnico di Torino, Corso Duca degli Abruzzi
24, TO, 10129, Torino, Italy, elena.ancona@gmail.com
\par\end{center}

\begin{center}
\textsuperscript{b}Department of Physics, New York City College of
Technology, The City University of New York, 300 Jay Street, Brooklyn
NY, 11201, New York City, USA, rkezerashvili@citytech.cuny.edu
\par\end{center}

\medskip{}

\begin{center}
\textbf{Abstract}
\par\end{center}

\noindent In this study we considered a solar sail coated with materials that undergo thermal desorption at a specific temperature, as a result of heating by solar radiation at a particular heliocentric distance. Three different scenarios, that only differ in the way the sail approaches the Sun, were analyzed and compared. In every case once the perihelion is reached, the sail coat undergoes thermal desorption. When the desorption process ends, the sail then escapes the Solar System having the conventional acceleration due to solar radiation pressure. Thermal desorption here comes as an additional source of solar sail acceleration beside traditional propulsion systems for extrasolar space exploration. The compared scenarios are the following: i. Hohmann transfer plus thermal desorption. In this scenario the sail would be carried as a payload to the perihelion with a conventional propulsion system by an Hohmann transfer from Earth's orbit to an orbit very close to the Sun (almost at 0.1 AU) and then be deployed there. ii. Elliptical transfer plus Slingshot plus thermal desorption. In this scenario the transfer occurs from Earth's orbit to Jupiter's orbit. A Jupiter's fly-by leads to the orbit close to the Sun, where the sail is deployed. iii. Two stage acceleration of the solar sail through thermal desorption. The proposed sail has two coats of the materials that undergo thermal desorption at different temperatures depending on the heliocentric distance. The first desorption occurs at the Earth orbit and provides the thrust needed to propel the solar sail toward the Sun. The second desorption is equivalent to that of the other scenarios.

\smallskip{}


\medskip{}

\begin{multicols}{2}

\section{Introduction}

Solar sails are large sheets of low areal density material whose only
source of energy is the Sun photons flux. At least in theory, a solar
sail mission could be of unlimited duration, thanks to the ``ever-present
gentle push of sunlight'' \citep{SolarSails}. A remarkable advantage
of solar sails is that no propellent is needed. We focused on the
possibilities of an high-performance sail to escape the Solar System
at enormous speed by means of thermal desorption of its coatings.
The dynamic efficiency of a solar sail as a propulsion device increases
upon its approach to the Sun. In particular, a solar sail can generate
a high cruise speed if it is deployed as close to the Sun as possible,
based on its reliable thermo-electrical and thermo-optical properties,
so that the force due to the solar radiation pressure is maximized.
We suggest using space environmental effects such as solar radiation
heating to accelerate a conventional solar sail for an extrasolar
space exploration. We are considering a solar sail coated with materials
that undergo desorption at a particular temperature, as a result of
heating by solar radiation at a particular heliocentric distance.
In our approach, the perihelion of the solar sail orbits is determined
based on the temperature requirement for the solar sail materials
and possibly where a peak perihelion temperature is obtained. This
acceleration boosts the solar sail to its escape velocity. However,
afterwards it will continue the accelerated motion due to the solar
radiation pressure. Therefore the solar sail, in addition to the conventional
acceleration caused by the solar radiation pressure, will take advantage
of an acceleration due to desorption of the coated materials. 

Our work is organized in the following way: in Section 2 we describe
the thermal desorption process, Section 3 is devoted to considerations
of the temperature dependence of solar sail materials on the heliocentric
distance. The orbital mechanics of the three considered scenarios
is discussed in Section 4, and finally results are given in Section
5, followed by conclusions in Section 6.

\section{Thermal desorption}

Thermal desorption is a process of mass loss which dominates all other
similar processes above temperatures of 300\textendash 500 \textdegree C.
Some special heat-sensitive materials can undergo the transition from
the solid state phase into the gas phase at particular temperature.
By heating to temperatures of 800\textendash 1000 K a sail on which
surface there is a coat of embedded atoms or paint, their thermal
desorption can provide higher specific impulse than liquid rockets,
as experimentally shown in Ref. \citep{microwave}. The chemical process
consists in atoms, embedded in a substrate, that are liberated by
heating, thus providing an additional thrust. Desorption can attain
high specific impulse if low mass molecules or atoms are blown out
of a lattice of material at high temperature. This idea was proposed
in Ref. \citep{BenfordDesorp}, where beam-powered microwave pulse
from the source located on the Earth or on an orbit is suggested to
use for heating (from Earth or from orbit). However, the solar sail
is naturally heated through the absorption of solar radiation. This
temperature rising can be used for the thermal desorption of the coatings
of the solar sail that will provide an additional thrust. After the
coats sublime away, the sail can perform as a conventional solar sail.
Acceleration of sails by thermal desorption is not a new idea, nevertheless
the innovation is to apply this concept to the solar sail that naturally
gains temperature through the absorption of solar radiation \citep{Rom_desorp}.

\subsection{The chemical process}

An adatom\footnote{abbreviation for ``adsorbed atom''.} present on
a surface at low temperatures may remain almost indefinitely in that
state. As the temperature of the substrate increases, however, there
will come a point at which the thermal energy of the adsorbed species
is such that it may desorb from the surface into the gas phase.

The rate of desorption $R_{d}$ of an adsorbed species from a surface
can be expressed in the general form as:

\begin{equation}
R_{d}=k_{d}\left(N_{A}\right)^{q},\label{eq:ratede}
\end{equation}
where $N_{A}$ is the surface concentration of adsorbed species, $q$
is the kinetic order of desorption and $k_{d}$ is the rate constant
for desorption. The latter one is commonly described by an Arrhenius
type equation \citep{thedes}:

\begin{equation}
k_{d}=\nu_{0}\:\exp\left(-\frac{E_{A}}{k_{B}\mathcal{T}}\right),\label{eq:costde}
\end{equation}
where $\nu_{0}$ is the pre-exponential frequency factor and its typical
value is $10^{13}s^{-1}$, $E_{A}$ is the activation or liberation
energy that is usually less than $1\,eV$, and $k_{B}=1.38\cdot10^{-23}\,JK^{-1}$
is the Boltzmann constant.

The order of desorption $q$ mostly depends on the type of considered
reaction: usually it is a first order process if it involves atomic
or simple molecular desorption. Sometimes $\nu_{0}$ is also called
the ``attempt frequency'' at overcoming the barrier to desorption.
In the particular case of simple molecular adsorption, $\nu_{0}$
corresponds to the frequency of vibration of the bond between the
atom or molecule and substrate. This is because every time the bond
is stretched during the course of a vibrational cycle could be considered
as an attempt to break the bond and, hence, an attempt at desorption.
In general it was found that the pre-exponential frequency factor
can span a wide range of values from $10^{13}s^{-1}$ to $10^{21}s^{-1}$
depending on the vibrational degrees of freedom of the adsorbate \citep{thedes}.

From Eqs. (\ref{eq:ratede}) and (\ref{eq:costde}) the general expression
for the rate of desorption can be obtained, considering that it corresponds
to the time reduction of adatoms present on surface:

\begin{equation}
R_{d}=-\frac{dN_{A}}{dt}=\nu_{0}\:\left(N_{A}\right)^{q}\:\exp\left(-\frac{E_{A}}{k_{B}\mathcal{T}}\right).\label{eq:costde-1}
\end{equation}
The rate of mass loss under heating $dN_{A}/dt$ is the desorbed flux
in $atoms/m^{2}s$. Note that the exponential factor suggests that
the desorption will have a sudden onset after the surface gets warm
\citep{BenfordDesorp}. Equation (\ref{eq:costde-1}) can be formally
solved when temperature varies with time. As the activation energy
increases, the time to desorb gets longer, because the rate of desorption
decreases.

A relation between the activation energy and the temperature at which
desorption occurs can be derived from Eq. (\ref{eq:costde-1}) and
it is known as the Redhead equation \citep{thedes,red}:

\begin{equation}
E_{A}=R\mathcal{T}_{md}\left(\ln\frac{\nu_{0}\mathcal{T}_{md}}{R_{h}}-3.64\right).\label{eq:ea}
\end{equation}
In Eq. (\ref{eq:ea}) $R$ is the universal gas constant, $\mathcal{T}_{md}$
is the temperature at which the rate of desorption reaches its maximum,
and $R_{h}$ is the heating rate. Experimental values for $E_{A}$
for carbon dioxide and monoxide can be found in Ref. \citep{thedes}.

Anyhow, desorption is a process which never occurs in isolation. Before
an atom or molecule can desorb it must reach the surface in some way,
lose energy - in order to reach an approximate thermal equilibrium
- and then gain energy from the surface. The adsorbed atom is never
in complete thermal equilibrium with the surface, and models that
begin with a thermal equilibrium assumption frequently encounter problems
which can be traced back to the fact that the true equilibrium state
is one in which the atom has already desorbed and is far from the
surface \citep{athedes}. Despite this, the model is still accurate
for preliminary calculations.

\subsection{Thermal desorption as a propulsion mechanism}

The thermal desorption was suggested as a propulsion method in \citep{microwave,Reducetime,DesorpAss}.
The original idea of Benford and Benford was to use a microwave beam
to heat a solar sail until its surface coat sublimes or desorbs. They
thought of two different scenarios: a ground-based microwave beamer
where the power-source is stationary on the ground, and alternatively
an orbital boosting method where the orbiting microwave beamer is
deployed behind a solar sail in the same initial circular orbit. In
this study instead, the suggestion is to use space environmental effects,
such as solar radiation heating, to accelerate a conventional coated
solar sail, as already proposed in Ref. \citep{Rom_desorp}.

Robert Forward first thought of microwave-driven sail as an alternative
to his laser sail concept \citep{ultraligh,laser}, but no experiments
were possible until light carbon appeared, because other materials
didn't allow liftoff under Earth gravity. As carbon sublimes instead
of melting, it can operate at very high temperature. James and Gregory
Benford took experiments on ultralight sails in carbon and found out
that photon pressure can account for 3 to 30\% of the observed acceleration,
while the remainder comes from desorption of embedded molecules \citep{microwave}.
Therefore, here is the extraordinary potential of this sort of propulsion
mechanism: if sapiently used, desorption could enhance thrust by many
orders of magnitude, shortening the escape time to weeks, instead
of years for conventional solar sails \citep{Reducetime}.

The acceleration from photon momentum $a_{P}$ produced by a power
$W$ on a thin film of mass $m$, area $A$ and reflectivity $\varrho$
is \citep{microwave}:

\begin{equation}
a_{P}=\left[\varrho+1\right]\frac{W}{m_{A}\cdot A\cdot c},\label{eq:phacc}
\end{equation}
where $c$ is the speed of light, $m_{A}=m/A$ is the mass per unit
area and corresponds to the solar sail's $\sigma$. Of the power incident
on the film, a fraction $\alpha W$ will be absorbed, which in steady
state must be radiated away from both sides of the film, according
to the Stefan-Boltzmann law:

\begin{equation}
\alpha W=2A\:\zeta\:\sigma_{SB}\:\mathcal{T}^{4},\label{eq:bolt}
\end{equation}
where $\zeta$ is the film emissivity that is assumed to be the same
for both sides of the film surface and $\sigma_{SB}=5.67\cdot10^{-8}W\,m^{-2}K^{-4}$
is the Stefan-Boltzmann constant. Substituting $W/A$ from (\ref{eq:bolt})
into (\ref{eq:phacc}) the acceleration clearly turns out to be strongly
limited by temperature $\mathcal{T}$:

\begin{equation}
a_{P}=\frac{2\sigma_{SB}}{c}\:\frac{\zeta\left[\varrho+1\right]}{\alpha}\:\frac{\mathcal{T}^{4}}{m_{A}},\label{eq:apot}
\end{equation}
where $2\sigma_{SB}/c=3.78\cdot10^{-16}J\,m^{-3}K^{-4}$. But there
is also another mechanism which should be considered: acceleration
due to the sublimation of the coating material mass. In fact, due
to mass ejected from the material in one direction, a force rises
in the opposite one, which determines the acceleration due to thermal
desorption,

\begin{equation}
a_{D}=\frac{v_{th}}{m}\frac{dm}{dt},\label{eq:ades}
\end{equation}
where $v_{th}$ is the thermal speed of the evaporated material. This
acceleration can vastly exceed the one of photon pressure, if the
temperature is high enough. For instance, $a_{D}/a_{P}\gg4.5\cdot10^{4}$
for molecular hydrogen at $\mathcal{T}=1000\,K$ as pointed in Ref.
\citep{BenfordDesorp}.

\begin{figure}[H]
\noindent \begin{centering}
\includegraphics[width=4.2cm]{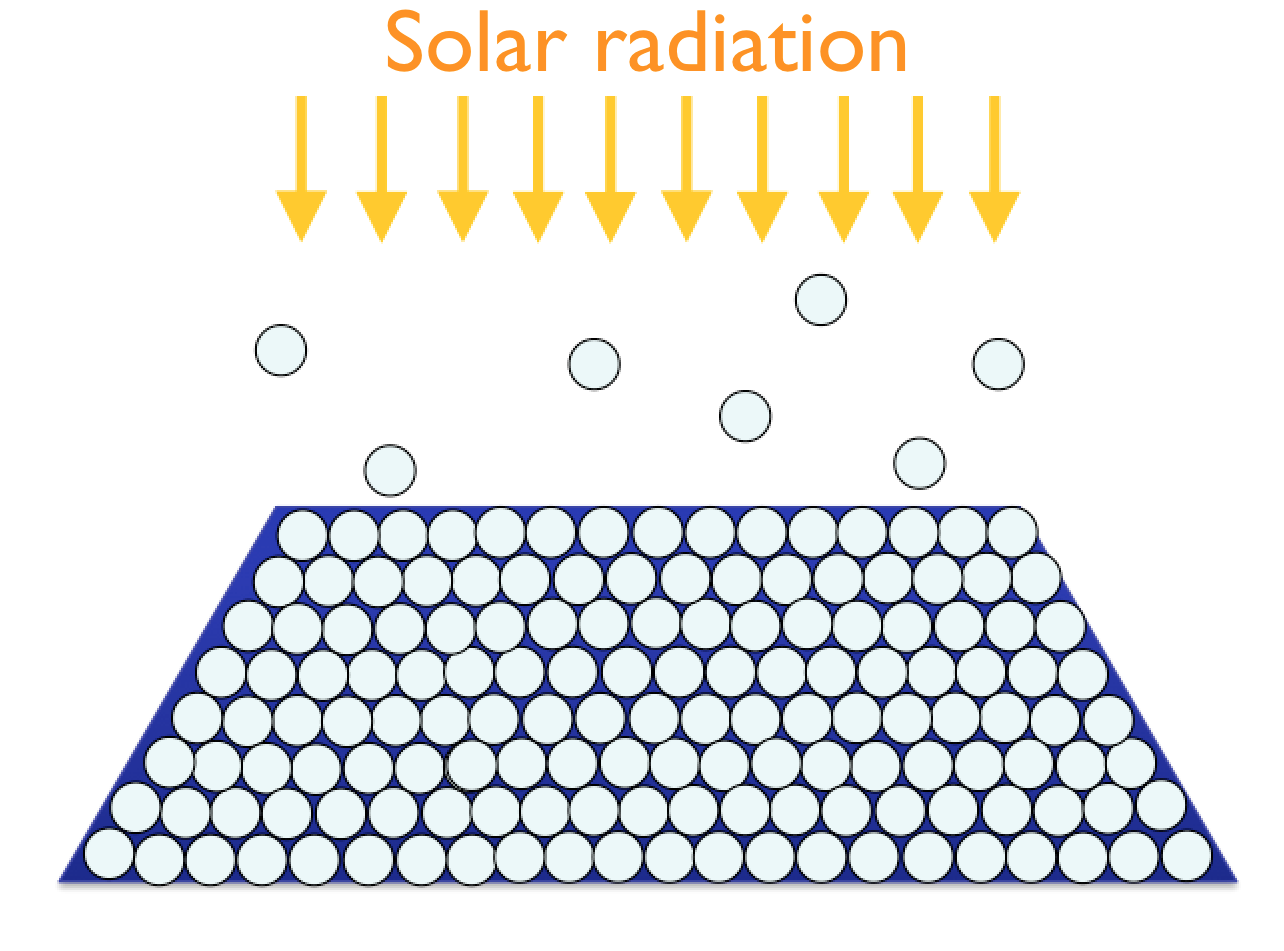}\includegraphics[width=4cm]{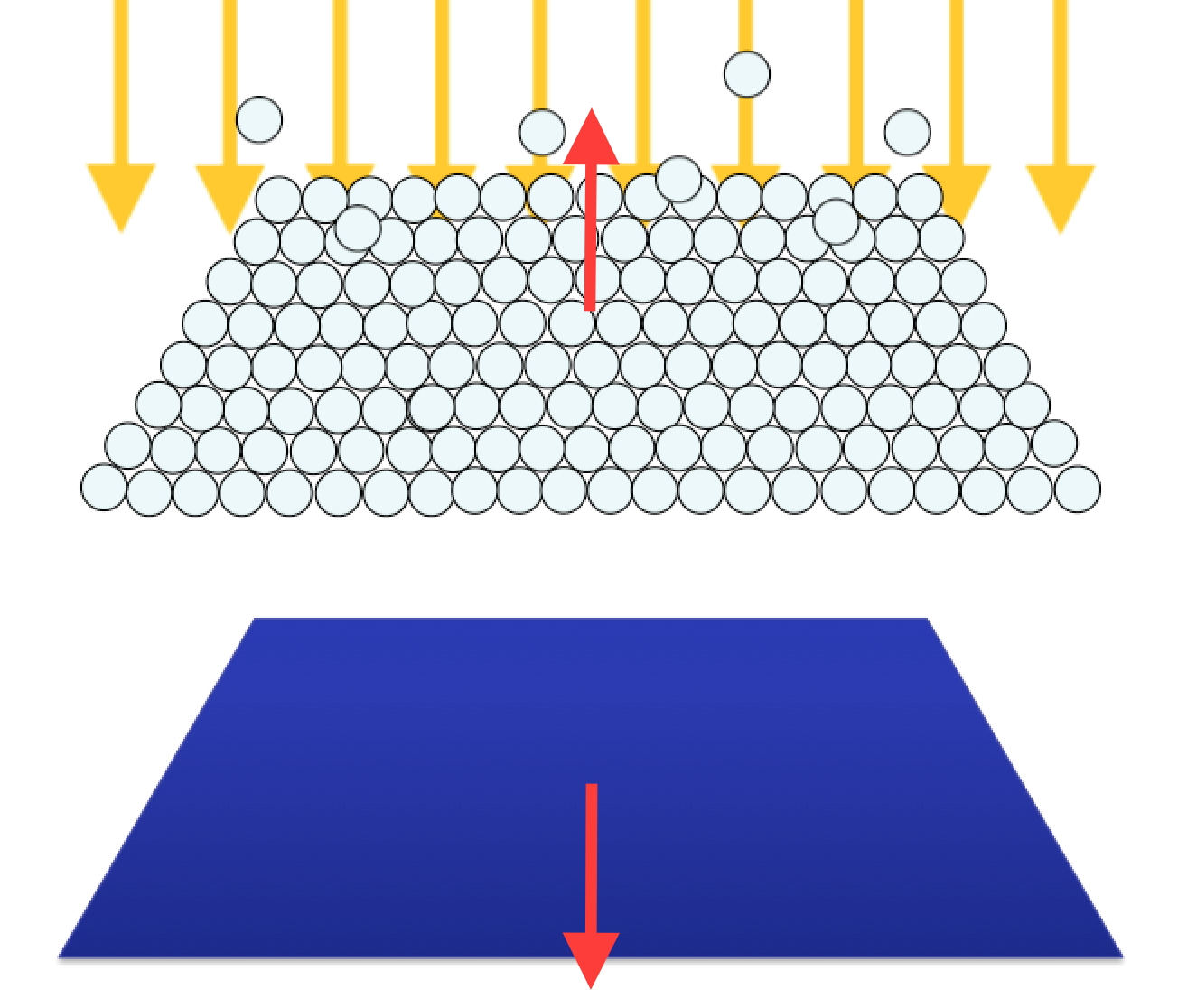}
\par\end{centering}

\caption{\selectlanguage{american}%
Thermal desorption propulsion mechanism.\selectlanguage{english}%
}

\label{desproc}
\end{figure}

Consider a sail coated with a material that undergoes desorption for
a particular temperature, corresponding to a specific heliocentric
distance. As the sail gets closer to the Sun, its temperature increases,
and some atoms of the coatings leave the surface. Once the right temperature
is reached, the material desorbs, as shown in Fig. \ref{desproc}.
If $N$ particles, molecules or atoms, of mass $m_{p}$ leave the
sail's surface at velocity $v_{th}$, as follows from the law of conservation
of total momentum, the sail of mass $m$ will move in the opposite
direction with velocity $v$, that is $v=\left(N\,m_{p}\,v_{th}\right)/m$. 

In order to find out the velocity of the sail, it is necessary to
evaluate the thermal speed $v_{th}$ of the evaporated particles.
The Maxwell speed distribution allows to calculate $v_{th}$ as the
mean thermal velocity:

\begin{equation}
v_{th}=\sqrt{\frac{8k_{B}\mathcal{T}}{\pi m_{p}}}.\label{eq:meanv}
\end{equation}
Assuming that the particles behave as ideal gas, and that they are
identical but still distinguishable, the Maxwell velocity distribution
function can be applied \citep{kittherm}:

\begin{equation}
\mathfrak{p}\left(v_{th}\right)=4\pi\left(\frac{m_{p}}{2\pi k_{B}\mathcal{T}}\right)^{3/2}v_{th}^{\;\;2}\;\exp\left(-\frac{m_{p}v_{th}^{\;\;2}}{2k_{B}\mathcal{T}}\right),\label{eq:pvth}
\end{equation}
where $\mathfrak{p}\left(v_{th}\right)$ is the probability that a
particle has its speed in the range $dv_{th}$ at $v_{th}$. Let $N\,\mathfrak{p}\left(v_{th}\right)dv_{th}$
be the number of atoms with speed in the range $dv_{th}$ at $v_{th}$,
so with velocity magnitude between $v_{th}$ and $v_{th}+dv_{th}$.
With the assumption that all the particles leave the substrate in
a direction orthogonal to the surface, their momentum is $\left[N\,\mathfrak{p}\left(v_{th}\right)dv_{th}\right]\,m_{p}\,v_{th}$.
Using the latter expression and Eq. (\ref{eq:pvth}) the total momentum
can be obtained by integrating it over $v_{th}=0$ to infinity:

\begin{equation}
p=\sqrt{8\pi\,N\,m_{p}\,\mathcal{T}}.\label{eq:p8}
\end{equation}

From Eq. (\ref{eq:p8}) follows that the momentum of thrust is proportional
to the square root of the temperature. Since the acceleration of the
solar sail is also proportional to the momentum,

\begin{equation}
a_{D}=\frac{\sqrt{8\pi\,N\,m_{p}\,\mathcal{T}}}{m\,\,t},\label{eq:adeso}
\end{equation}
it follows that even acceleration is proportional to $\sqrt{\mathcal{T}}$
and to the inverse of time. This result can be compared to the photon
pressure acceleration $a_{P}$ in Eq. (\ref{eq:apot}), which was
strongly limited by temperature, being proportional to $\mathcal{T}^{4}$.

\subsection{Thermal desorption for a sail coating}

Now consider a sail coated with a material that undergoes thermal
desorption. The proposed coating is in carbon, as suggested in Ref.
\citep{BenfordDesorp}. The acceleration due to thermal desorption
is given by Eq. (\ref{eq:adeso}), where $N$ is the number of particles
desorbed, $m_{p}$ is the mass of desorbed atoms or molecules, $\mathcal{T}$
is the temperature for which desorption occurs, $m$ is the sail mass
and $t$ is the time needed for the physical process of thermal desorption
to be completed. 

Once the material is known $m_{p}$ is determined, and if also the
thickness and the area of the sail, $H$ and $A$ respectively, are
defined, one can find the number of atoms desorbed. In fact, being
$\mathfrak{d}$ the sail's material density and $V=H\cdot A$ its
volume, as $N=m/m_{p}$ one gets:

\begin{equation}
N=\frac{\mathfrak{d}\:H\:A}{m_{p}}.
\end{equation}
Then the number of atoms desorbed per unit area during the time, $dN_{A}/dt$,
is obtained from (\ref{eq:costde-1}) if the activation energy $E_{A}$
is known (see Ref. \citep{c60,graphite}). If the sail area is defined,
$N_{A}$ is fixed, so the required desorption time can be found \citep{actpreexp,desact}.
However this approach would require very accurate information about
the materials characteristics, and the activation energy varies in
a wide range and would require ad hoc experiments to have reliable
data. For this reason another approach was used: as suggested in Benford's
paper \citep{BenfordDesorp} a nominal rate of mass loss was chosen,
$dm/dt=1\:g/s$. For a given sail area $A$ and thickness $H$ or
loading factor $\sigma=m/A$, the mass of the sail is fixed. For each
heliocentric distance $r_{P}$ (to which corresponds a defined temperature
$\mathcal{T}$) one can evaluate the mean thermal velocity using Eq.
(\ref{eq:meanv}) that by means of (\ref{eq:ades}) provides the acceleration
due to desorption. Given a sail coating thickness $H_{c}$, the density
of the material and the area of the sail, the total mass required
for the desorption is obtained, and then the desorption time dividing
the mass over the rate of mass loss.

\subsubsection*{Solution of the problem }

Considering an heliocentric distance $r_{P}=0.3\:AU$ (to which corresponds
$\mathcal{T}=737\:K$ as will be shown in the next Section) the mean
thermal velocity for Carbon is $1168\:m/s$ and the acceleration due
to desorption $0.39\:m/s^{2}$. If the sail coating thickness is $H_{c}=5\cdot10^{-6}\:m$
the total mass required for the desorption is $M_{c}=34\:kg$, and
the desorption time $34005\:s$ (almost 9 hours). Because the mass
of the coating is not negligible with respect to the sail mass, and
the desorption time is also relevant, the formula $v=v_{0}+a\cdot t$
doesn't apply. In fact, this analysis cannot neglect the variation
of mass during the acceleration time. Being the mass of the coated
sail $m=\sigma A+M_{c}$, the force on the sail is:

\begin{equation}
F=\frac{d}{dt}\left(\sigma A+M_{c}\right)v=\left(\sigma A+M_{c}\right)\frac{dv}{dt}+\frac{dM_{c}}{dt}v,
\end{equation}
whereas the force due to thermal desorption is: 

\begin{equation}
F=\frac{dM_{c}}{dt}v_{th}.
\end{equation}
By comparing these two equations one obtains a differential equation
in $v$:

\begin{equation}
\frac{dv}{dt}+\frac{1}{\left(\sigma A+M_{c}\right)}\frac{dM_{c}}{dt}v-\frac{1}{\left(\sigma A+M_{c}\right)}\frac{dM_{c}}{dt}v_{th}=0,\label{eq:diffeqv}
\end{equation}
that, by introducing the following notations,

\[
E=\frac{1}{\left(\sigma A+M_{c}\right)}\frac{dM_{c}}{dt},\quad G=\frac{1}{\left(\sigma A+M_{c}\right)}\frac{dM_{c}}{dt}v_{th},
\]
can be re-written as:

\begin{equation}
\frac{dv}{dt}+E\cdot v-G=0.\label{eq:diffeqv-1}
\end{equation}
One can solve Eq. (\ref{eq:diffeqv-1}) and obtain the expression
for the velocity as function of time:

\begin{equation}
v(t)=\left(v_{0}-\frac{G}{E}\right)e^{-Et}+\frac{G}{E}.
\end{equation}

\section{Temperature dependence on heliocentric distance}

In the previous section the acceleration of the sail has been written
as function of temperature, by means of Eqs. (\ref{eq:apot}) and
(\ref{eq:adeso}). However, temperature of the solar sail surface
depends on the heliocentric distance \citep{Temp_IAC}. The aim of
this section is to determine how the temperature depends on heliocentric
distance. Such analysis can be done based on the law of conservation
of energy by considering the temperature dependence of the reflectivity,
absorptivity and emissivity of solar sail materials \citep{materials}. 

Due to the interaction with the surface of the solar sail, the solar
electromagnetic radiation can be reflected, absorbed or transmitted.
Therefore, one can write:

\begin{equation}
\rho\left(\lambda,\mathit{\mathcal{T}}\right)+\alpha\left(\lambda,\mathcal{T}\right)+\tau\left(\lambda,\mathcal{T}\right)=1,
\end{equation}
where $\rho$, $\alpha$ and $\tau$ are the radiative or optical
properties of the solar sail material: spectral (as they depend on
wavelength) hemispherical (as they are not directional) reflectivity,
absorptivity and transmissivity, respectively \citep{heatmass}.

To evaluate the pressure exerted on the sail, it is necessary to consider
the momentum transported not by a single photon, but by a flux of
photons. Based on the inverse square law the energy flux $\phi$ (also
said solar irradiance) depends on the distance $r$ of the body from
the Sun as: 

\begin{equation}
\phi\left[\frac{W}{m^{2}}\right]=\phi_{E}\left(\frac{R_{E}}{r}\right)^{2},\label{eq:flux}
\end{equation}
where $R_{E}=1\:AU$ is the Sun-Earth distance and $\phi_{E}=L_{S}/\left(4\pi R_{E}^{2}\right)=1346\:\:W/m^{2}$
is the Solar irradiance at Earth distance, defined through the solar
luminosity $L_{S}$.

Considering the sail pitch angle $\vartheta$, the incident solar
energy flux $\phi$ in (\ref{eq:flux}) is the following:

\begin{equation}
\phi=\phi_{E}\cdot cos\vartheta\left(\frac{R_{E}}{r}\right)^{2}.\label{eq:incfl}
\end{equation}

The corresponding absorbed energy flux will be:

\begin{equation}
\phi_{a}=\left[1-\rho\left(\mathcal{T}\right)-\tau\left(\mathcal{T}\right)\right]\phi,
\end{equation}
where the dependence on wavelength has been removed computing the
spectral average, so considering the total hemispherical optical coefficients.

Once that energy has been absorbed, it can also be emitted from the
surface, as a secondary process. From the Stefan-Boltzmann's law,
the rate of energy emitted from a surface at a certain temperature
is proportional to the fourth power of the temperature. As a sail
is usually composed by various films of different materials, in general
its front and back sides will have different emissivity, $\zeta_{F}$
and $\zeta_{B}$, respectively. Hence, the emitted energy flux, from
front and back surfaces is:

\begin{equation}
\phi_{e}=\left[\zeta_{F}\left(\mathcal{T}\right)+\zeta_{B}\left(\mathcal{T}\right)\right]\cdot\sigma_{SB}\cdot\mathcal{T}^{4}.
\end{equation}

From the law of conservation of energy the solar sail will be in thermal
equilibrium if the total absorbed energy equals the total emitted
energy. By applying the equality $\phi_{e}=\phi_{a}$, the so known
Kirchhoff\textquoteright s law, and considering that from Eq. (\ref{eq:incfl})
the solar energy flux has an inverse square variation with the heliocentric
distance $r$, one can find the dependence of the sail temperature
on the heliocentric distance as
\begin{equation}
\mathcal{T}^{4}=\left\{ \frac{\left[1-\rho-\tau\right]\cdot\phi_{E}\cdot cos\vartheta\cdot R_{E}^{2}}{\left[\zeta_{F}+\zeta_{B}\right]\cdot\sigma_{SB}}\right\} r^{-2}.
\end{equation}

If the front and back sides have almost equal emissivity ($\zeta\simeq\zeta_{F}\approx\zeta_{B}$),
the latter equation gives:

\begin{equation}
\mathcal{T}=\left\{ \frac{\left[1-\rho-\tau\right]\cdot\phi_{E}\cdot cos\vartheta\cdot R_{E}^{2}}{2\cdot\zeta\cdot\sigma_{SB}}\right\} ^{1/4}r^{-1/2}.\label{eq:temp12}
\end{equation}

From the analysis of Eq. (\ref{eq:temp12}) it results that $\mathcal{T}\varpropto r^{-1/2}$,
but this is true only within the standard approach that considers
the optical coefficients to have constant values and no dependence
on temperature \citep{SolarSailing}. Experimental given data are
$\rho=0.88$ and $\zeta=0.03$ for Aluminum \citep{radheat}. However
recent studies show that reflectivity and emissivity also depend on
temperature \citep{materials}. Although reflectivity dependence on
temperature can be neglected, the same doesn't apply for emissivity,
which is directly proportional to the temperature as suggested by
Parker and Abbott in Ref. \citep{emiss}. The expression they found
for the total hemispherical emissivity for metals is the following:

\begin{equation}
\begin{array}{c}
\zeta\left(\mathcal{T}\right)=7.66\,\sqrt{\frac{\mathcal{T}}{\chi\left(\mathcal{T}\right)}}+\left[10+8.99\ln\left(\frac{\mathcal{T}}{\chi\left(\mathcal{T}\right)}\right)\right]\cdot\\
\cdot\left(\frac{\mathcal{T}}{\chi\left(\mathcal{T}\right)}\right)-17.5\cdot\left(\frac{\mathcal{T}}{\chi\left(\mathcal{T}\right)}\right)^{3/2},
\end{array}\label{eq:zeta-2}
\end{equation}
where $\chi\left(\mathcal{T}\right)$ is the electrical conductivity
in $\Omega^{-1}\:m^{-1}$, which is defined as the inverse of resistivity
$\widetilde{\rho}\left(\mathcal{T}\right)$: 

\begin{equation}
\chi\left(\mathcal{T}\right)\equiv\frac{1}{\widetilde{\rho}\left(\mathcal{T}\right)}.\label{eq:chi1}
\end{equation}
It has been shown with experimental results \citep{radheat} that
a very good approximation can be obtained considering only the first
term of the Eq. (\ref{eq:zeta-2}), that leads to:

\begin{equation}
\zeta\left(\mathcal{T}\right)=7.66\,\sqrt{\frac{\mathcal{T}}{\chi\left(\mathcal{T}\right)}}.\label{eq:zeta}
\end{equation}

For Aluminum the coefficient in (\ref{eq:zeta}) is $7.52\:K^{-1/2}\:\Omega^{-1/2}\:m^{-1/2}$.
The electrical resistivity of most materials changes with temperature.
Usually a linear approximation is used:

\begin{equation}
\widetilde{\rho}\left(\mathcal{T}\right)=\widetilde{\rho}_{0}\left[1+t{}_{coeff}\left(\mathcal{T}-\mathcal{T}_{0}\right)\right]\simeq\frac{\widetilde{\rho}_{0}}{\mathcal{T}_{0}}\,\mathcal{T},\label{eq:chi2}
\end{equation}
where $t{}_{coeff}\simeq1/273\:K^{-1}$ is called the temperature
coefficient of resistivity, $\mathcal{T}_{0}$ is a fixed reference
temperature (commonly room temperature) and $\widetilde{\rho}_{0}$
is the resistivity at temperature $\mathcal{T}_{0}$. Values for electrical
conductivity, resistivity and temperature coefficient of various materials
can be found in literature. Considering a reference temperature of
$\mathcal{T}_{0}=293\:K$, for Aluminum $\widetilde{\rho}_{0}=2.82\cdot10^{-8}\,\Omega\cdot m$,
$\chi{}_{0}=3.50\cdot10^{7}\,S/m$ and $t{}_{coeff}=0.0039\,K^{-1}$.

As it follows from (\ref{eq:chi1}) and (\ref{eq:chi2}), $\chi\left(\mathcal{T}\right)$
is almost inversely proportional to the temperature, this means that
emissivity $\zeta\left(\mathcal{T}\right)\propto\mathcal{T}$: 

\begin{equation}
\zeta\left(\mathcal{T}\right)=7.52\,\sqrt{\mathcal{T}\cdot\widetilde{\rho}\left(\mathcal{T}\right)}=7.52\,\mathcal{T}\sqrt{\frac{\widetilde{\rho}_{0}}{\mathcal{T}_{0}}}.\label{eq:zeta-1}
\end{equation}

By introducing this analysis in Eq. (\ref{eq:temp12}) the dependence
of temperature on the heliocentric distance varies as

\begin{equation}
\mathcal{T}=\left\{ \frac{\left[1-\rho-\tau\right]\cdot\phi_{E}\cdot cos\vartheta\cdot R_{E}^{2}}{2\cdot\left[7.52\,{\textstyle {\displaystyle \sqrt{\frac{\widetilde{\rho}_{0}}{\mathcal{T}_{0}}}}}\right]\cdot\sigma_{SB}}\right\} ^{1/5}r^{-2/5}.\label{eq:dipeft}
\end{equation}

\begin{figure*}[t!]

\begin{minipage}[c]{\textwidth} \centering 

\selectlanguage{american}%
\includegraphics[width=15cm]{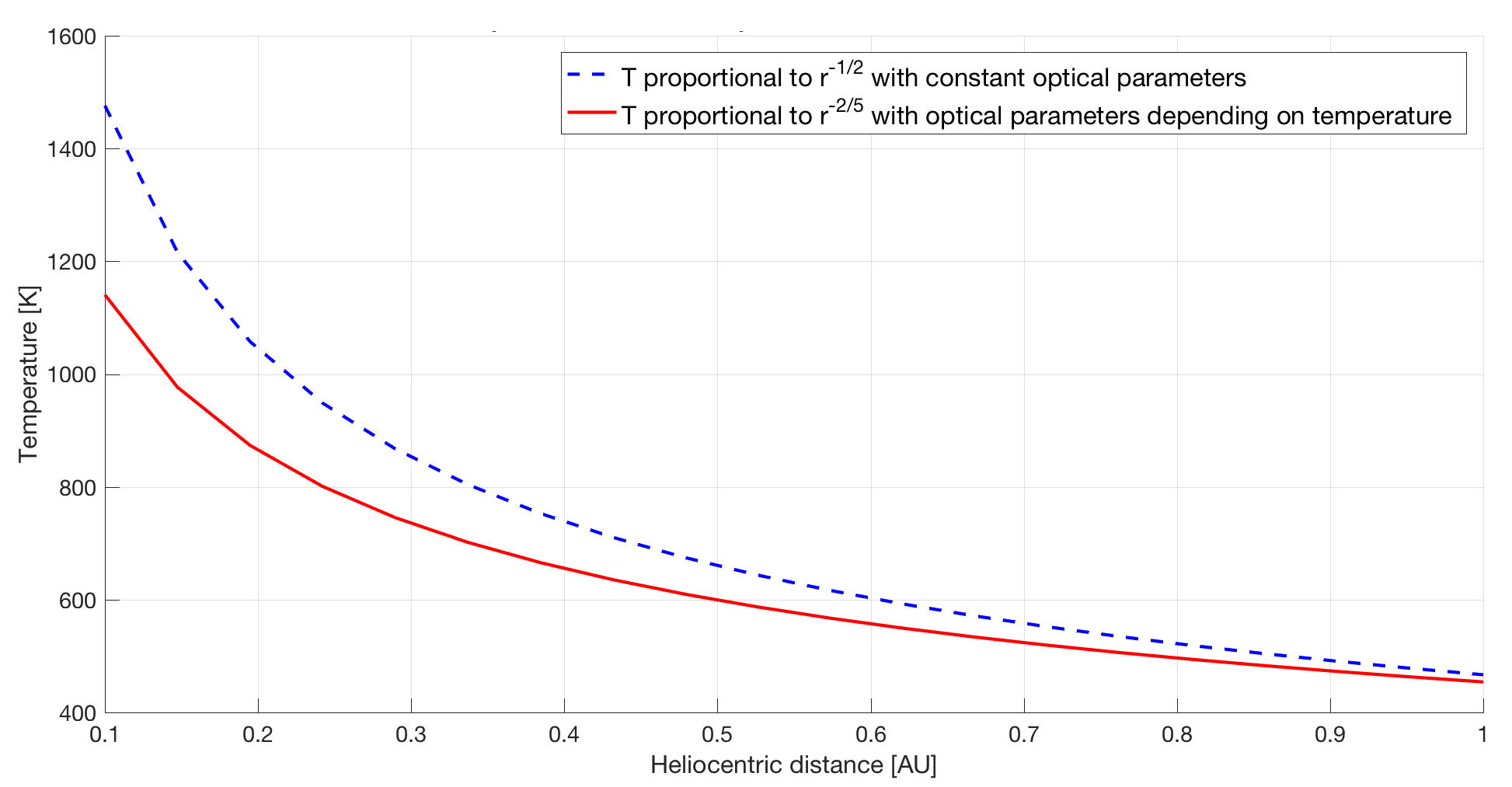}

\selectlanguage{english}%
\end{minipage}

\captionof{figure}{Dependence of sail temperature on the heliocentric distance.}

\label{heldist}

\end{figure*}

\bigskip{}

Dependence of the sail temperature on the heliocentric distance for
a case of constant emissivity and, conversely, when emissivity of
the solar sail material depends on temperature is shown in Fig. \ref{heldist},
and listed in Table \ref{temp}. It is clear that, considering the
temperature dependence of emissivity (solid curve), the sail temperature
increases more slowly than in the case of constant emissivity (dashed
curve), as the sail approaches the Sun \citep{Rom_desorp}; this leads
to the conclusion that the sail would be able to get closer to the
Sun without degradation of its optical properties. 
Figure \ref{heldist} compares Eq. (\ref{eq:temp12}), the blue line,
to Eq. (\ref{eq:dipeft}), the red line. In our calculations for both
the Equations the sail pitch angle was considered to be $\vartheta=0\text{\textdegree}$
and the transmissivity $\tau$ was neglected, as experimental data
confirm it is a very small fraction (almost 2\%) of the incoming solar
energy flux.

\medskip{}

\captionof{table}{Dependence of sail temperature on the heliocentric distance.}

\begin{singlespace}
\noindent \begin{center}
\begin{tabular}{rrr}
\hline 
Distance {[}AU{]} & T {[}K{]} $\mathcal{T}\propto r^{-1/2}$ & T {[}K{]} $\mathcal{T}\propto r^{-2/5}$\tabularnewline
 & ($\zeta=cost$) & ($\zeta=f\left(\mathcal{T}\right)$)\tabularnewline
\hline 
$0.1$ & 1476.1 & 1140.6\tabularnewline
$0.2$ & 1043.8 & 864.4\tabularnewline
$0.3$ & 852.2 & 735.0\tabularnewline
$0.4$ & 738.1 & 655.1\tabularnewline
$0.5$ & 660.1 & 599.2\tabularnewline
$0.6$ & 602.6 & 557.0\tabularnewline
$0.7$ & 557.9 & 523.7\tabularnewline
$0.8$ & 521.9 & 496.5\tabularnewline
$0.9$ & 492.0 & 473.6\tabularnewline
$1$ & 466.8 & 454.1\tabularnewline
\hline 
\end{tabular}
\par\end{center}
\end{singlespace}

\label{temp}

In our scenarios analysis the considered temperature will be the one
deriving from the more accurate analysis with the thermal variation
of optical properties of the sail material.

\begin{figure}[H]
\noindent \begin{centering}
\includegraphics[width=8cm]{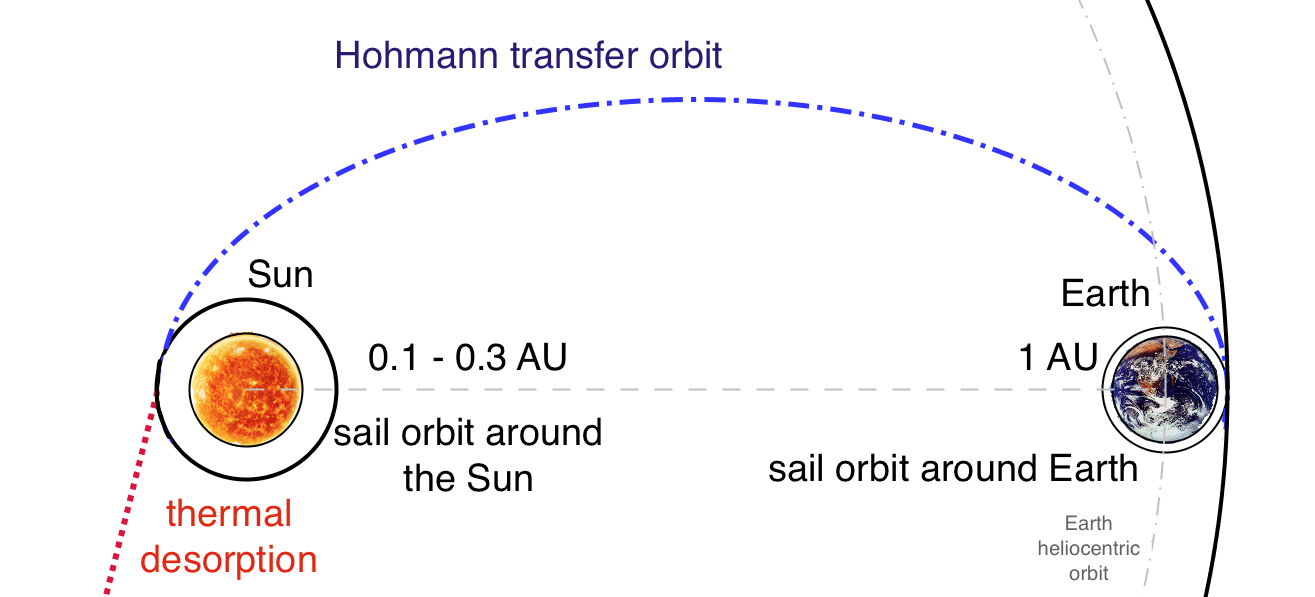}
\par\end{centering}

\caption{\selectlanguage{american}%
First scenario, Hohmann transfer plus thermal desorption acceleration
(figure not to scale).\selectlanguage{english}%
}

\selectlanguage{american}%
\label{1stscen}\selectlanguage{english}%
\end{figure}

\section{Proposed scenarios}

This section will focus on the orbital mechanics of a solar sail accelerated
by thermal desorption of coatings and different scenarios that could
be considered. It is very important to remark that in order to have
an effective desorption, high temperatures are required, so every
scenario takes advantage of a passage extremely close to the Sun.

\begin{figure*}[t!]

\begin{minipage}[c]{\textwidth} \centering 

\noindent \begin{center}
\includegraphics[width=17cm]{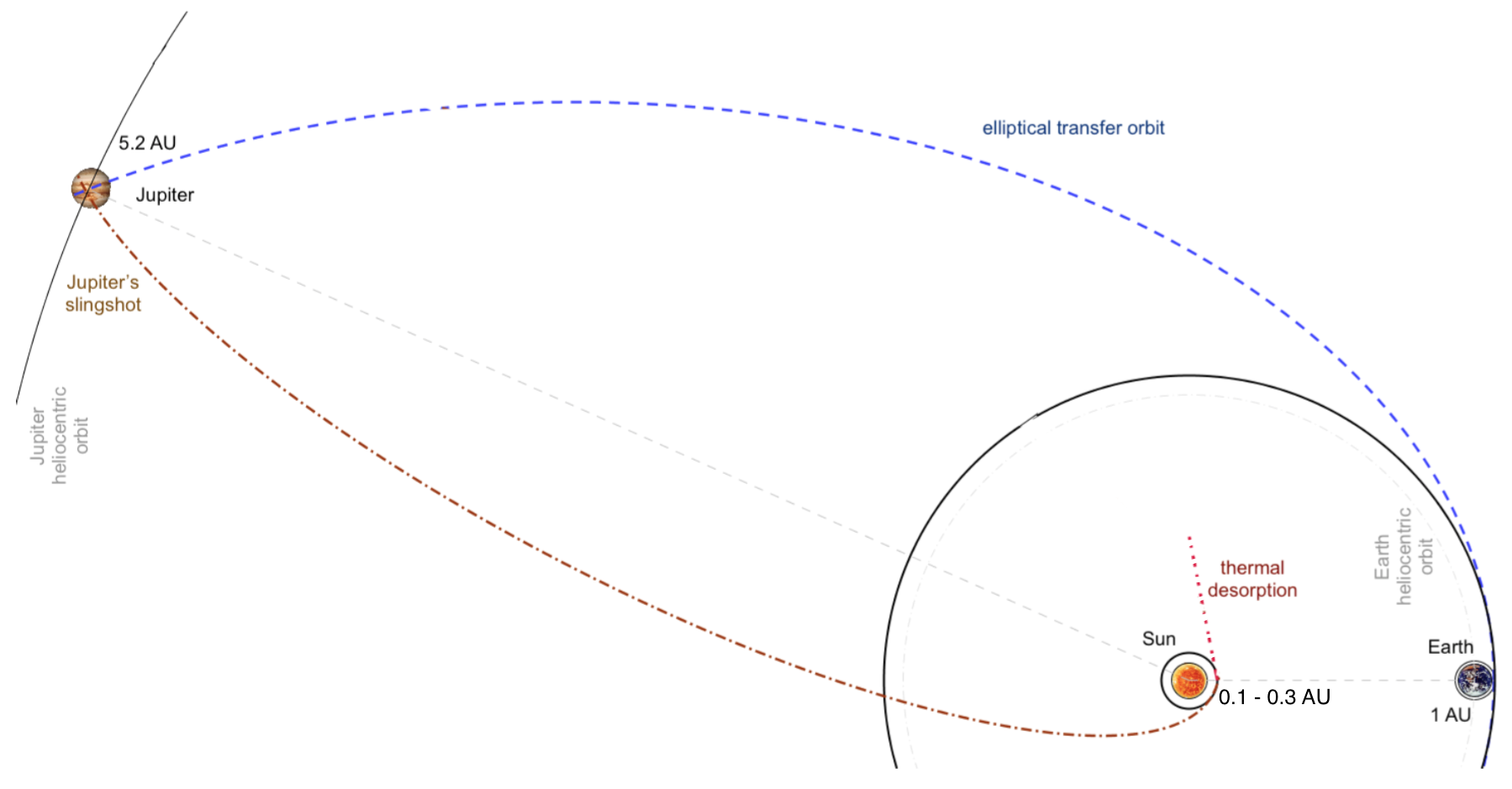}
\par\end{center}

\end{minipage}

\captionof{figure}{Second scenario, elliptical transfer plus slingshot plus thermal desorption (figure not to scale).}

\label{2ndscen}

\end{figure*}

For every scenario the transfer orbit to the Sun is in dash-dot line,
the thermal desorption from the perihelion distance is in dotted line.
In solid line the orbits of the planets or the sail. For the second
scenario, the transfer orbit from Earth to Jupiter is in dashed line.

\subsection{Hohmann transfer plus thermal desorption}

This is the simplest scenario that can be esteemed. It involves a
Hohmann transfer from Earth's orbit to an orbit very close to the
Sun, the perihelion of latter varies from 0.3 AU to 0.1 AU. The proposed
sail would be carried to the perihelion with a conventional propulsion
system and then be deployed there. The solar sail has one coat of
the material that undergoes desorption at the temperature reached
at the perihelion of the transfer orbit. The sail then escapes the
Solar System. The schematic diagram for the first scenario is shown
in Fig. \ref{1stscen}.

\subsection{Elliptical transfer plus Slingshot plus thermal desorption acceleration}

In this scenario a generic elliptical transfer is performed from Earth's
orbit to Jupiter's orbit. Then a Jupiter's fly-by leads to the orbit
close to the Sun, where the sail is deployed and thermal desorption
accelerates it to the escape velocity. As in the previous case, the
sail has one coat of the material that undergo desorption at the temperature
reached at the perihelion. The corresponding schematic diagram for
this scenario is shown in Fig. \ref{2ndscen}.

\begin{figure}[H]
\noindent \begin{centering}
\includegraphics[width=5cm]{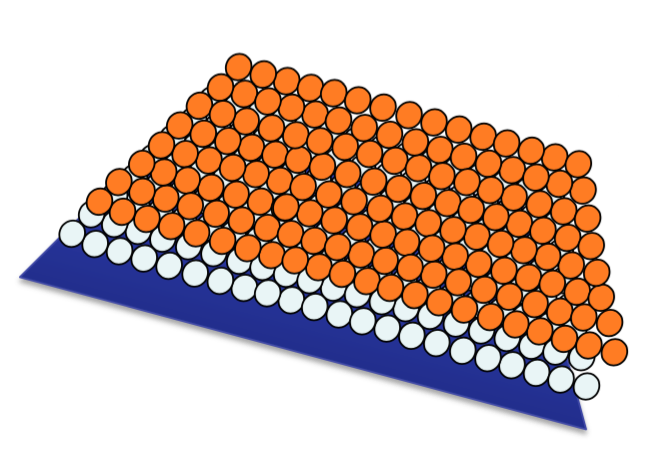}
\par\end{centering}

\caption{\selectlanguage{american}%
Solar sail with two coatings of different materials proposed for the
third scenario.\selectlanguage{english}%
}
\end{figure}

\subsection{Two stage acceleration of the solar sail through thermal desorption}

\begin{figure}[H]
\noindent \begin{centering}
\includegraphics[width=8cm]{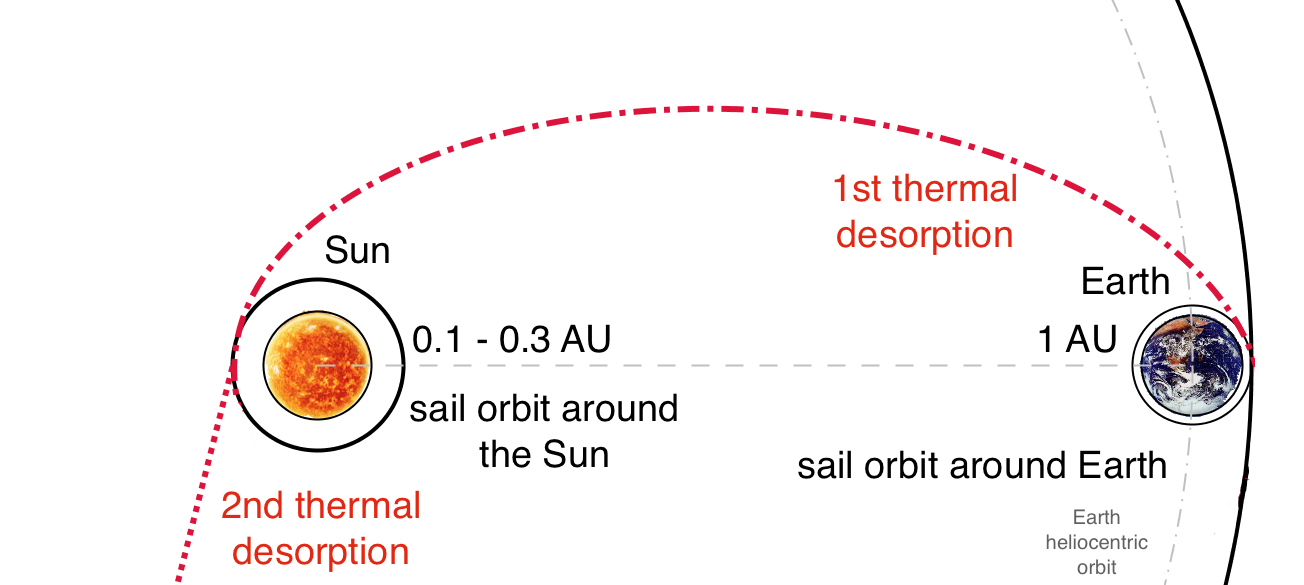}
\par\end{centering}

\caption{Third scenario, two stage acceleration through thermal desorption
(figure not to scale).}

\label{3rdscen}
\end{figure}

The proposed sail has two coats of the materials that undergo desorption
at different solar sail temperatures depending on the heliocentric
distance. The first desorption occurs at the Earth orbit and provides
the thrust needed to propel the solar sail toward the Sun. When the
solar sail approaches the Sun, its temperature increases, and the
second coat undergoes desorption at the perihelion of the heliocentric
escape orbit. This provides a second thrust and boosts the solar sail
to its escape velocity. The schematic diagram for this scenario is
shown in Fig. \ref{3rdscen}. Note that in this scenario the sail
is deployed from the beginning, so it will actually approach the Sun
obeying to the conventional solar sail orbital dynamics, with a typical
spiral logarithmic trajectory. It is important to remark that, although
we are considering an high-performance solar sail (so with lightness
number $\beta>1$) for our mission, the $\beta$ that has to be considered,
in order to evaluate the spiral logarithmic trajectory, is the one
of the inner coating, as the outer would have left the sail at the
beginning of the heliocentric transfer. The coating material must
have good properties for desorption, so we accept a low lightness
number for the first part of the mission (resulting basically in a
longer time required to get to the perihelion).

These three scenarios are analyzed and compared. As a preliminary
analysis the planets are considered point-like. In the heliocentric
reference frame, assuming by approximation that the Earth's orbit
is almost circular, with radius $r_{2}$, the sail has to be transferred
to an inner orbit closer to the Sun, of radius $r_{1}<r_{2}$, in
order to escape the Solar System. The transfer between these two coplanar
circular orbits is different for every scenario.\bigskip{}

\section{Results}

For the sake of brevity we will skip the orbital dynamics calculations,
that are nothing but the application of well-known conventional formulas
\foreignlanguage{american}{\citep{Bate}}, and go directly to the
results that show the positive contribution of thermal desorption.
In Table \ref{scentab} we present the results obtained for all the
proposed scenarios. Solving Eq. (\ref{eq:diffeqv}) the final velocity
of the sail after the desorption time $v_{D+}$ can be determined.
By contrast, $v_{D-}$ is the velocity that the sail has when it reaches
the perihelion, before desorption occurs, and it comes from the orbital
dynamics of: i) a conventional spacecraft for the first two scenarios;
ii) a solar sail with constant pitch angle for the third scenario.

\captionof{table}{Scenarios comparison. For each scenario and each radius of perihelion, the corresponding speed before desorption, speed after desorption, cruise speed and distance covered over years are given.}

\selectlanguage{american}%
\begin{singlespace}
\noindent \begin{center}
\begin{tabular}{llrrr}
\hline 
\multicolumn{5}{c}{\selectlanguage{english}%
1\textsuperscript{st }scenario\selectlanguage{american}%
}\tabularnewline
\hline 
$r_{P}$ & $\left[AU\right]$ & 0.3 & 0.2 & 0.1\tabularnewline
$v_{D-}$ & $\left[km/s\right]$ & 67.39 & 85.91 & 126.89\tabularnewline
$v_{D+}$ & $\left[km/s\right]$ & 167.16 & 213.43 & 315.81\tabularnewline
$v_{cruise}$ & $\left[km/s\right]$ & 170.65 & 217.54 & 321.43\tabularnewline
$AU_{y}$ & $\left[AU/year\right]$ & 35.9 & 45.8 & 67.7\tabularnewline
\hline 
\end{tabular}
\par\end{center}

\noindent \begin{center}
\begin{tabular}{llrrr}
\hline 
\multicolumn{5}{c}{\selectlanguage{english}%
2\textsuperscript{nd} scenario\selectlanguage{american}%
}\tabularnewline
\hline 
$r_{P}$ & $\left[AU\right]$ & 0.3 & 0.2 & 0.1\tabularnewline
$v_{D-}$ & $\left[km/s\right]$ & 73.24 & 91.28 & 129.11\tabularnewline
$v_{D+}$ & $\left[km/s\right]$ & 181.82 & 226.90 & 321.44\tabularnewline
$v_{cruise}$ & $\left[km/s\right]$ & 185.04 & 230.76 & 326.90\tabularnewline
$AU_{y}$ & $\left[AU/year\right]$ & 39.0 & 48.7 & 68.9\tabularnewline
\hline 
\end{tabular}
\par\end{center}

\noindent \begin{center}
\begin{tabular}{llrrr}
\hline 
\multicolumn{5}{c}{\selectlanguage{english}%
3\textsuperscript{rd} scenario\selectlanguage{american}%
}\tabularnewline
\hline 
$r_{P}$ & $\left[AU\right]$ & 0.3 & 0.2 & 0.1\tabularnewline
$v_{D-}$ & $\left[km/s\right]$ & 54.33 & 64.77 & 91.60\tabularnewline
$v_{D+}$ & $\left[km/s\right]$ & 134.42 & 160.44 & 227.41\tabularnewline
$v_{cruise}$ & $\left[km/s\right]$ & 138.74 & 165.87 & 235.07\tabularnewline
$AU_{y}$ & $\left[AU/year\right]$ & 29.2 & 34.9 & 49.5\tabularnewline
\hline 
\end{tabular}
\par\end{center}
\end{singlespace}

\selectlanguage{english}%
\label{scentab}

The final cruise speed $v_{cruise}$ is obtained from the application
of the law of conservation of energy, and from it comes the distance
travelled per year. First of all, one can notice that when reducing
the heliocentric distance of perihelion, the speed increases for all
the three scenarios. By analyzing the results it is clear that the
higher velocity is reached at the perihelion, the greater will be
the contribution of desorption and consequently the cruise speed of
the solar sail. In terms of cruise speed and distance covered per
year, the best scenario is always the second one, that takes advantage
of a profitable planetary flyby. In fact, for the heliocentric distance
of $0.1\:AU$ the cruise speed obtained is extremely high: almost
$327\:km/s$. A solar sail traveling at this speed would cover a distance
of $69\:AU$ per year. This means that it would take only about 8
years to reach the Sun gravity focus, at $550\:AU$. Note that for
a conventional solar sail typical values of cruise speed correspond
to $15\:AU/year$.

It is also evident that the first and second scenario have very similar
values of speed. However, the second scenario would be preferable
for a relevant reason, that cannot be deducted from the table below,
and it is that the $\Delta V$ required for the first scenario is
much more expansive than the one for the scenario with Jupiter flyby.
In particular, for the most demanding heliocentric distance to reach,
$0.1\:AU$, the first scenario requires a $\Delta V=12.53\:km/s$,
whereas for the second scenario $\Delta V=7.05\:km/s$. The only disadvantage
of the second scenario is that it would require more time: the Hohmann
transfer to $0.1\:AU$ would need just 73 days, instead the flyby
scenario requires 558 days only to get to Jupiter, and it is necessary
to add to this value also the days required to get back to $0.1\:AU$.
From the point of view of time required, the second scenario is, with
no doubt, the most critical one. In fact even the spiral logarithmic
trajectory of the sail of the third scenario would be faster: 474
days required with $\beta=0.1$ and only 107 days with $\beta=0.4$,
respectively. 

\begin{figure}[H]
\noindent \begin{centering}
\includegraphics[width=8.4cm]{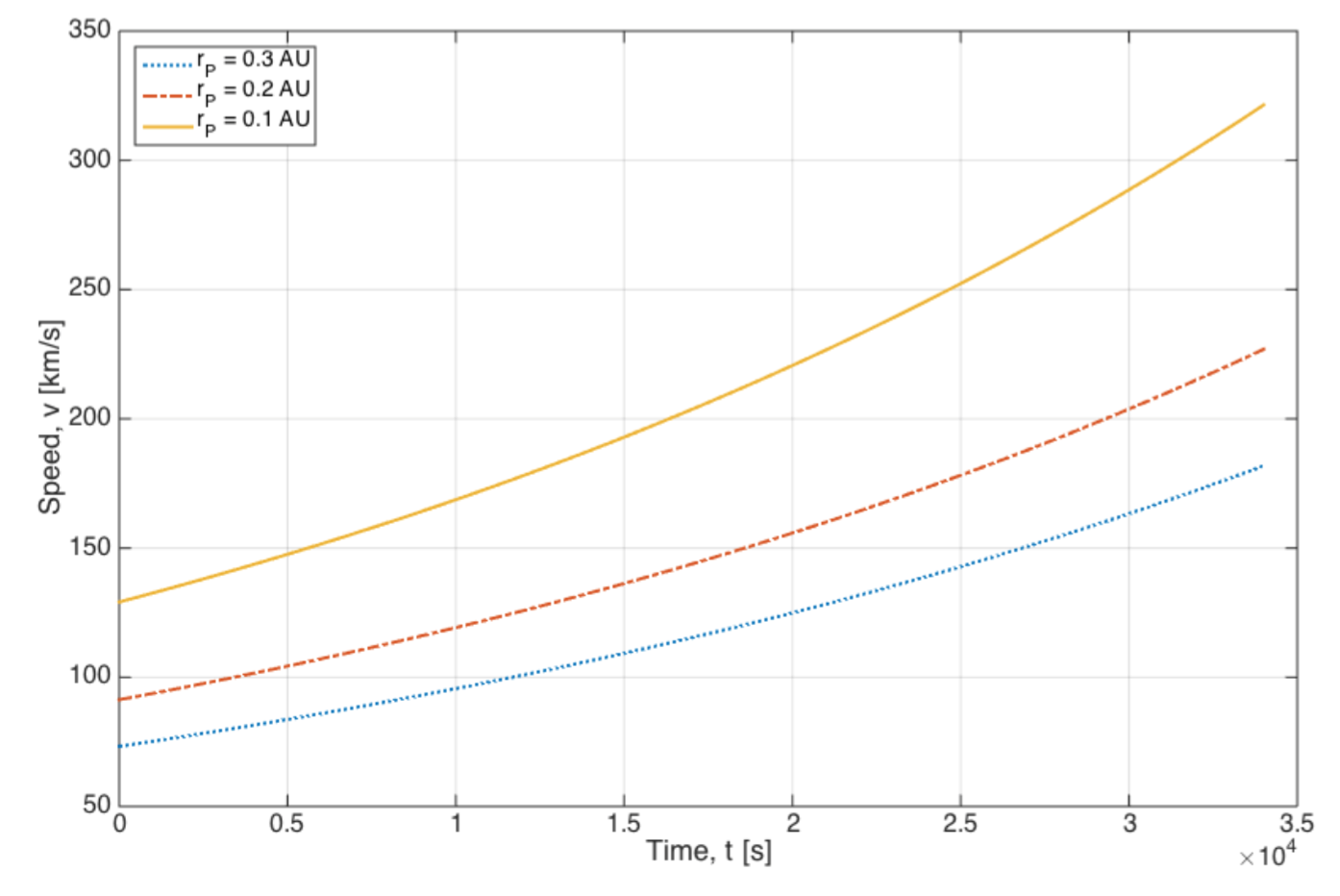}
\par\end{centering}

\caption{Gain in velocity due to thermal desorption for the second scenario.}

\label{des}
\end{figure}
We considered $\beta=0.1$ because, as previously explained, the $\Delta V$
required would dramatically grow for $\beta=0.4$. In fact, $\Delta V=2.5\:km/s$
for the third scenario with $\beta=0.1$, whereas $\Delta V=10.9\:km/s$
with $\beta=0.4$. Let us mention that the $\Delta V$ required for
the third scenario is given starting from a LEO, so it cannot be compared
to the previous two scenarios, where the launch was included in the
$\Delta V$ budget calculations.

To conclude, the performance of a two steps acceleration of the solar
sail by thermal desorption are lower than the ones than it is possible
to achieve with a Jupiter flyby, in terms of velocity reached and
distance covered per year. However, this configuration could be advantageous
in terms of times and $\Delta V$ required.

Let us put in evidence the benefits of thermal desorption: if sapiently
driven, this phenomenon can enhance a lot the speed of the solar sail,
as one can see by simply comparing the values of $v_{D+}$ and $v_{D-}$
for all the scenarios. The acceleration due to desorption is of the
order of $m/s^{2}$, whereas that of photon pressure is a few $mm/s^{2}$,
for this reason the photon pressure acceleration is neglected during
the short desorption time. Fig. \ref{des} shows the gain in velocity
due to desorption for the best scenario. Eq. \ref{eq:adeso} shows
that the acceleration of desorption is proportional to the temperature,
and this is also confirmed in the graph.

\section{Conclusions}

Let us summarize our work and come to conclusions. The concept on
which all the study is based is that for extrasolar space exploration
it might be very convenient to take advantage of space environmental
effects, such as solar radiation heating. A solar sail could be accelerated
if coated by materials that undergo thermal desorption at a particular
temperature. Thermal desorption is a physical process of mass loss
which dominates all other similar processes and it can provide additional
thrust as heating liberates atoms, embedded on the surface of a solar
sail.

We considered a solar sail coated with one or two materials that undergo
thermal desorption at a specific temperature, as a result of heating
by solar radiation at a particular heliocentric distance, and focused
on the orbital dynamics of three scenarios that only differ in the
way the sail approaches the Sun. In every case once the perihelion
is reached, the sail coat undergoes thermal desorption. When the desorption
process ends, the sail then escapes the solar system having the conventional
acceleration due to solar radiation pressure. Our study analyzed and
compared three different scenarios in which thermal desorption comes
beside traditional propulsion systems for extrasolar space exploration.

The first scenario consists of a Hohmann transfer plus thermal desorption.
In this scenario the sail is carried as a payload to the perihelion
with a conventional propulsion system by a Hohmann transfer from Earth's
orbit to an orbit very close to the Sun, the perihelion of latter
varies from 0.3 AU to 0.1 AU, and then is deployed there. The second
proposed scenario consists of an elliptical transfer plus slingshot
plus thermal desorption. In this scenario the transfer occurs from
Earth's orbit to Jupiter's orbit and then a Jupiter's fly-by leads
to the orbit close to the Sun, where the sail is deployed. The last
and third scenario consists of a two stage acceleration of the solar
sail through thermal desorption. The proposed sail has two coats of
the materials that undergo thermal desorption at different temperatures
depending on the heliocentric distance. The first desorption occurs
at the Earth orbit and provides the thrust needed to propel the solar
sail toward the Sun. The second desorption is the same as in the other
scenarios.

The comparison of the scenarios shows that the second one is the best
in terms of cruise speed and distance covered per year. In fact, for
the heliocentric distance of $0.1\:AU$ the cruise speed obtained
is almost $327\:km/s$, that leads to $69\:AU$ travelled per year.
For all the scenarios, however, the great advantage of thermal desorption
is clear and evident. 

Our study is a preliminary design of a solar sail mission considering
the innovative contribution of thermal desorption of one or more coatings.
At this stage many assumptions were made in order to simplify the
solution of the problem. Eventually the study could be improved and
refined. For example, for the third scenario a better result could
be obtained considering the sail pitch angle $\vartheta$ not constant
but variable, with a numerical methods optimization. Also, let us
mention that the perturbations of other bodies were not taken into
account. It could be done with the help of numerical analysis, starting
from the patched conics method and propagating the orbit with some
tools such as STK. Another relevant assumption is that the Sun was
always considered as point-like body. But when approaching the Sun
very close, the inverse square law results to be inadequate and requires
consideration of the finite size of the Sun. This topic is treated
in \citep{SolarSailing}.

The natural prosecution of this work would be a detailed research
on materials, in order to find out their characteristics, performances
and desorption temperature suitable for solar sailing. For what concerns
materials, in fact, the research could be extended widely. Carbon
nanotubes coatings have been suggested for solar sails \Citep{Rom_desorp,E_1},
but there could be better solutions for the specific mission proposed.
This would require also laboratory experiments to evaluate activation
energies and all the other properties involved in the chemical thermal
desorption process.

\section*{Acknowledgements }

This research was supported by PSC CUNY Grant: award \# 68298-0046. 

\bibliographystyle{unsrtnat}

%\bibliographystyle{unsrtnat}
%\addcontentsline{toc}{section}{\refname}\bibliography{SolarSailBiblio}

\end{multicols}{}
\end{document}